\documentclass[12pt,psfig]{article}
\usepackage{amssymb}
\usepackage{amsmath,amsthm}
\usepackage{amsfonts}
\usepackage{graphicx}
\usepackage{graphics}
\usepackage{indentfirst}
\numberwithin{equation}{section}

\usepackage{hyperref}

\begin{document}
\begin{center}\Large\textbf{LREE of an Unstable Dressed-Dynamical
D$p$-brane: Superstring Calculations}
\end{center}
\vspace{0.75cm}
\begin{center}{\large Shirin Teymourtashlou and \large Davoud
Kamani}\end{center}
\begin{center}
\textsl{\small{Department of Physics, Amirkabir University of
Technology (Tehran Polytechnic) \\
P.O.Box: 15875-4413, Tehran, Iran \\
e-mails: sh.teymourtash@aut.ac.ir , kamani@aut.ac.ir \\}}
\end{center}
\vspace{0.5cm}

\begin{abstract}

We obtain the left-right entanglement entropy (LREE) for a
D$p$-brane with tangential motion in the presence of a $U(1)$
gauge potential, the Kalb-Ramond field and an open string
tachyon field. Thus, at first we extract the R\'{e}nyi entropy
and then by taking a special limit of it we  acquire the
entanglement entropy. We shall investigate 
the behavior of the LREE 
under the tachyon condensation phenomenon. We observe that 
the deformation of the LREE, through
this process, reveals the collapse of the brane.
Besides, we examine the second law of thermodynamics
for the LREE under tachyon condensation, and we extract the
imposed constraints. Note that our calculations 
will be in the context of the type IIA/IIB superstring theories.

\end{abstract}

{\it PACS numbers}: 11.25.Uv; 11.25.-w

\textsl{Keywords}:
Background fields; Tangential dynamics; Boundary state;
Interaction amplitude; Left-right entanglement entropy;
Tachyon condensation; Thermodynamics.

\newpage
\section{Introduction}

Entanglement is one of the essential features of
quantum mechanics. It correlates subsystems of
a composite quantum system such that 
the quantum state of each subsystem
cannot be described independently of  
the quantum states of the other subsystems.
For a composite quantum system in a pure state,
an applicable tool for measuring the 
entanglement between the subsystems
is the entanglement entropy. This adequate quantity has been
extensively studied, for example, in the context of the AdS/CFT
there have been evidences for relations between the 
entanglement entropy and gravity \cite{1}, \cite{2}. 
In addition, a connection between the 
black hole entropy and entanglement entropy
has been shown \cite{3}, \cite{4}. Besides, the 
entanglement entropy has been employed in condensed matter 
and the many-body 
quantum systems \cite{5}, \cite{6}, \cite{7}.

Traditionally, the entangled subsystems are separated
geometrically which leads to a separation in the 
Hilbert space. However, in this paper 
the division of the subsystems occurs only 
in the Hilbert space. That is,
the left- and right-moving modes of closed superstring 
form the subspaces. The entropy of the entanglement
between the left- and right-moving 
modes is called the left-right
entanglement entropy (LREE) \cite{8}-\cite{12}.

The crucial role of the D-branes in string theory has
been highly remarked in the literature. Various areas
such as the AdS/CFT, black holes and
string phenomenology prominently depend on the D-brane dynamics.
Since the boundary states accurately elaborate all properties
of the associated D-branes, they have been commonly used
in the brane analysis \cite{13}-\cite{26}.
In this paper, we shall investigate  
the LREE of a special D$p$-brane via the 
associated boundary state to it.

The early works were done by L. P. Zayas and N. Quiroz. They
studied the LREE, associated with a one-dimensional
boundary state, in a 2D CFT \cite{8}. Then, they developed
their analysis to a bare-static D$p$-brane \cite{9}.
Their works motivated us to extend the LREE calculations
for a dressed-dynamical D$p$-brane \cite{27}, 
and, afterward, for an unstable
dressed-dynamical D$p$-brane \cite{28}. Our papers
have been written in the context of the bosonic string theory.

The current study will be in the context of the 
type IIA/IIB superstring theories. Therefore,
we shall derive the LREE of a D$p$-brane with 
the tangential rotation and tangential 
linear motion, in the presence of an 
internal $U(1)$ gauge potential, the Kalb-Ramond field
and an open string tachyon field. 
In fact, there are some evidences for connection
between the entanglement entropy and 
black hole entropy \cite{3, 4}. Hence, the LREE of 
our configuration may find a relation with
the Bekenstein-Hawking entropy of the 
rotating-charged black holes.

Note that adding the open string tachyon to
a D-brane gives rise to instability.
Consequently, after condensing the tachyon, 
the brane looses its dimension, and one receives a 
lower-dimensional unstable brane
\cite{29}-\cite{37}. Accordingly, presence of the open string
tachyon on our brane enforces
the brane to collapse. We shall examine 
the behavior of the LREE under this experience. 
We shall see that the LREE of the D$p$-brane is decomposed 
to the LREE of a D$(p-1)$-brane and an extra 
contribution which might be associated with the emitted 
closed superstrings via the brane collapse.  
In comparison with the bosonic case \cite{28}
a D-brane in the superstring theory is 
more stable than its counterpart in the 
bosonic string theory. Moreover, we shall see that the 
thermal entropy of the setup exactly 
is equivalent to its LREE.
Because of the resemblance between 
the thermal and entanglement
entropies \cite{38}-\cite{42}, we investigate the
second law of thermodynamics for the LREE through
tachyon condensation process. We find that 
the survival of the second law imposes some conditions 
on the parameters of the configuration.

The paper is organized as follows.
In Sec. 2, we introduce the boundary state, corresponding to 
the dressed-dynamical D$p$-brane, and
subsequently the interaction amplitude between two
such D$p$-branes will be written. 
This amplitude will be required for computing 
the left-right R\'{e}nyi entropy. In Sec. 3, we obtain 
the LREE of our setup. In addition, we derive the thermodynamic
entropy, which is equivalent to our LREE. In Sec. 4, the 
evolution of the LREE under the tachyon condensation
phenomenon will be investigated. 
The second law of thermodynamics on the change of 
the LREE will be examined.
Section 5 will be devoted to the results and conclusions.

\section{The interaction amplitudes via the boundary states}

\subsection{The bosonic part of the boundary state}

At first, we obtain the bosonic part of the boundary state,
associated with a dynamical D$p$-brane
in the presence of the Kalb-Ramond field $B_{\mu \nu}$,
the $U(1)$ gauge potential $A_{\alpha}(X)$ and the open
string tachyon field $T(X)$. Therefore, we start with the
following string action
\begin{eqnarray}
S=&-&\frac{1}{4\pi\alpha'}\int_{\Sigma} {\rm d}^2\sigma
\left(\sqrt{-g} g^{ab}G_{\mu\nu} \partial_{a}X^\mu
\partial_b X^{\nu} +\varepsilon^{ab}
B_{\mu\nu}\partial_a X^{\mu}
\partial_b X^{\nu}\right)
\nonumber\\[10pt]
&+&\frac{1}{2\pi\alpha'}\int_{\partial\Sigma}
{\rm d}\sigma\left(A_{\alpha}
\partial_{\sigma}X^{\alpha}
+\omega_{\alpha\beta}J^{\alpha\beta}_{\tau}
+T(X^\alpha) \right),
\label{eq:2.1}
\end{eqnarray}
where the sets
$\{\sigma^a|a=0,1\}$ and
$\{x^\alpha|\alpha=0, 1,\cdots,p\}$ represent the
worldsheet coordinates and the parallel directions
to the brane worldvolume, respectively.
The set $\{x^i|i=p+1,\cdots,9\}$ will be used for the
perpendicular directions to the brane worldvolume.
We shall take the flat worldsheet 
and spacetime with the signature
$G_{\mu \nu}=\eta_{\mu\nu}=\rm{diag}(-1,1,...,1)$. Besides,
we apply a constant antisymmetric tensor $B_{\mu \nu}$.
The spacetime angular velocity  $\omega_{\alpha \beta}$
includes the tangential rotation and 
tangential linear motion of the brane,
and the angular momentum density is denoted by
$J^{\alpha\beta}_\tau=X^\alpha \partial_\tau X^\beta
-X^\beta \partial_\tau X^\alpha$.
For the gauge potential we use the profitable gauge
$A_\alpha=-\frac{1}{2}F_{\alpha \beta} X^\beta$ with the
constant field strength $F_{\alpha\beta}$, and the
tachyon profile is adopted as
$T=\frac{1}{2} U_{\alpha \beta}X^\alpha X^\beta$, with
$U_{\alpha\beta}$ as a constant and symmetric matrix.
We should mention that due to the presence of the
various fields on the brane worldvolume the Lorentz symmetry
has been manifestly lost. This clarifies  
that the tangential dynamics
along the brane worldvolume is meaningful.

By varying the action with respect to $X^\mu$
we find the equation of motion and the
flowing equations for the boundary state
\begin{eqnarray}
&~&\left( \Delta_{\alpha\beta} \partial_\tau X^\beta
+\mathcal{F}_{\alpha\beta}
\partial_{\sigma} X^\beta+B_{\alpha i}
\partial_\sigma X^i+U_{\alpha\beta}
X^\beta \right)_{\tau=0} |B_x\rangle=0,
\nonumber\\[10pt]
&~&\left(X^i-y^i\right)_{\tau=0}|B_x\rangle=0, 
\label{eq:2.2}
\end{eqnarray}
where $\mathcal{F}_{\alpha\beta}\equiv B_{\alpha\beta}
- F_{\alpha\beta}$ and
$\Delta_{\alpha\beta}\equiv \eta_{\alpha\beta}
+4\omega_{\alpha\beta}$.
The parameters $\{y^i\}$ exhibit the
position of the brane. Applying the mode expansion of $X^\mu$,
we can conveniently express the above equations in terms of the
closed string oscillators
\begin{eqnarray}
&~&\left[ \left(\Delta_{\alpha\beta}
-\mathcal{F}_{\alpha\beta}+\frac{i}{2m}
U_{\alpha\beta}\right) \alpha^\beta_m
+ \left(\Delta_{\alpha\beta}+
\mathcal{F}_{\alpha\beta}-\frac{i}{2m}
U_{\alpha\beta}\right)
\tilde{\alpha}^\beta_{-m}\right]|B_x ^{(\rm osc)}\rangle=0,
\nonumber\\[10pt]
&~&\left(2\alpha'\Delta_{\alpha\beta}\;p^\beta
+U_{\alpha\beta}\;x^\beta\right)
|B_x ^{(0)}\rangle=0, 
\label{eq:2.3}
\end{eqnarray}
for the parallel directions to the brane worldvolume, and
\begin{eqnarray}
&~& (\alpha^i_m-\tilde{\alpha}^i_{-m})
|B_x ^{(\rm osc)}\rangle=0,
\nonumber\\
&~& (x^i-y^i)|B_x ^{(0)}\rangle=0,
\label{eq:2.4}
\end{eqnarray}
for the normal directions. Note that 
we applied the decomposition
$|B_x\rangle = |B_x ^{(\rm osc)}\rangle 
\otimes |B_x ^{(0)}\rangle$.

By employing
the coherent state method and quantum mechanical techniques,
specially the commutation relations 
among the string oscillators,
we receive
\begin{eqnarray}
|B_x ^{(0)}\rangle&=&\frac{T_p}{2\sqrt{\det(U/4\pi \alpha')}}
\int_{-\infty}^{\infty} \prod_{\alpha=0}^p
\exp\bigg[ i\alpha' \sum_{\beta \neq \alpha}
(U^{-1}\Delta+\Delta^{\rm T}\;
U^{-1})_{\alpha\beta}p^\alpha p^\beta
\nonumber\\[10pt]
&+&\frac{i \alpha'}{2}(U^{-1}\Delta
+\Delta^{\rm T}\; U^{-1})_{\alpha\alpha}
(p^\alpha)^2 \bigg]
|p^\alpha \rangle {\rm d}p^\alpha
\nonumber\\[10pt]
&\times& \prod_{i=p+1}^{9}
\left[ \delta(x^i-y^i) |p^i=0\rangle \right], 
\label{eq:2.5}
\\[10pt]
|B_x ^{(\rm osc)}\rangle &=&
\prod_{n=1}^{\infty} [-\det M_{(n)}]^{-1}
\exp\left[ -\sum_{m=1}^{\infty}
\left( \frac{1}{m} \alpha^{\mu}_{-m}
S_{(m)\mu\nu} \tilde{\alpha}^\nu _{-m} \right) \right]
|0\rangle_\alpha |0\rangle_{\tilde \alpha}, 
\label{eq:2.6}
\end{eqnarray}
where the brane tension is $T_p$,
and we defined
$S_{(m)\mu\nu}=(Q_{(m)\alpha\beta},-\delta_{ij})$,
in which
\begin{eqnarray}
Q_{(m)\alpha\beta}&\equiv&(M_{(m)}^{-1}
N_{(m)})_{\alpha\beta},
\nonumber\\
M_{(m)\alpha\beta}&=&\Delta_{\alpha\beta}
-\mathcal{F}_{\alpha\beta}
+\frac{i}{2m}U_{\alpha\beta},
\nonumber\\
N_{(m)\alpha\beta}&=&\Delta_{\alpha\beta}
+\mathcal{F}_{\alpha\beta}
-\frac{i}{2m}U_{\alpha\beta}.
\label{eq:2.7}
\end{eqnarray}
The prefactors of both parts of $|B_x\rangle$ originate
from the normalization of the disk partition function.
For more details see Refs. \cite{27, 28}.
In fact, the boundary state $|B_x \rangle$
is not normalizable, i.e., 
the inner product $\langle B_x|B_x \rangle$ is divergent. 
In Sec. 3, we will introduce the regularization factor 
$e^{-\epsilon H}/\sqrt{\mathcal N_B}$, with a
finite correlation length $\epsilon$ and a
suitable normalization factor ${\mathcal N_B}$, 
to fix this problem. 

The first equation in Eq. (\ref{eq:2.3}) 
tells us that applying the
coherent state method on the set
$\{\alpha^\alpha_m ,{\tilde \alpha}^\alpha_{-m}|m
\in \mathbb{N}\}$ gives a boundary state with the matrix
$Q_{(m)\alpha\beta}$, while employing that method on the set
$\{{\tilde \alpha}^\alpha_m ,\alpha^\alpha_{-m}|m
\in \mathbb{N}\}$ yields a boundary state which includes
the matrix
$\left( \left[Q_{(-m)}^{-1}\right]^{\dagger}
\right)_{\alpha\beta}$. Equality of the resultant states
imposes the following conditions on the parameters of the setup
\begin{eqnarray}
&~& \Delta\; U=U\; \Delta^{\rm T},
\nonumber\\
&~&\Delta\; \mathcal{F}=\mathcal{F}\;
\Delta^{\rm T}. 
\label{eq:2.8}
\end{eqnarray}

The conformal ghosts also contribute to the bosonic part
of the boundary state as in the following
\begin{equation}
|B_{\rm gh}\rangle=\exp \left[ \sum_{n=1}^\infty
(c_{-n}\tilde b_{-n}-b_{-n} \tilde c_{-n})\right]
\frac{c_0+\tilde c_0}{2}\; |q=1\rangle\; |\tilde q=1\rangle.
\label{eq:2.9}
\end{equation}

\subsection{The fermionic part of the boundary state}

The unstable D$p$-brane in our setup carries an open string
tachyonic mode. In fact, survival of the open string tachyon
after the GSO projection requires our D-brane
to be a non-BPS D-brane with the wrong
dimension, i.e., odd (even) dimension in the type IIA (IIB)
theories. Therefore, the brane 
worldvolume does not couple to the
R-R form fields of the type II theories, and hence it cannot
carry any R-R charges. The corresponding boundary state to
a non-BPS brane merely possesses the NS-NS sector
$|B\rangle=|B\rangle_{\rm NS-NS}$ \cite{37, 43, 44}, also see
Ref. \cite{9}. Thus, in this paper we apply only the NS-NS 
sector of the type II theories.

Due to the worldsheet supersymmetry, we can perform
the following replacements on the bosonic boundary state
equations (\ref{eq:2.2}) to obtain their fermionic
counterparts
\begin{eqnarray}
\partial_+ X^\mu (\sigma,\tau) \; &\to& \;
-i\eta \psi^{\mu} _+ (\tau+\sigma),
\nonumber\\
\partial_- X^\mu (\sigma,\tau) \; &\to&
\; - \psi^{\mu} _- (\tau-\sigma), 
\label{eq:2.10}
\end{eqnarray}
in which $\partial_\pm=(\partial_\tau \pm \partial_\sigma)/2$.
The factor $\eta=\pm 1$ originates from the
boundary conditions on the fermionic coordinates and will
be used in the GSO projection on the boundary
state. Because of the presence of the tachyonic field,
a replacement for  $X^\mu$ is also needed. Employing the above
replacements and the mode expansions for $\psi^\mu_{\pm}$, 
we acquire
\begin{equation}
X^\mu(\sigma, \tau) \to \sum_t \frac{1}{2t}
\left( i\psi^\mu _t e^{-2it(\sigma-\tau)}+
\eta \tilde \psi ^\mu _t e^{-2it(\sigma+\tau)} \right),
\label{eq:2.11}
\end{equation}
where the index ``$t$'' is half-integer for
the NS-NS sector.

Applying the replacements (\ref{eq:2.10}) and (\ref{eq:2.11})
into Eqs. (\ref{eq:2.2}), and also using the mode expansion
of $\psi^\mu _{\pm}$, we obtain
\begin{eqnarray}
&~&\left[  \left(\Delta_{\alpha \beta}
-\mathcal F_{\alpha \beta}+\frac{i}{2t} U_{\alpha \beta}\right)
\psi^\beta _t -i\eta
\left( \Delta_{\alpha \beta}
+\mathcal F_{\alpha \beta}-\frac{i}{2t} U_{\alpha \beta}\right)
\tilde \psi^\beta _{-t}\right] |B _{\psi} ,\eta \rangle=0,
\nonumber\\
&~&( \psi_t ^i+i\eta \tilde \psi_{-t} ^i)
|B _\psi ,\eta \rangle=0.
\label{eq:2.12}
\end{eqnarray}
Eqs. (\ref{eq:2.12}) can be combined as
\begin{eqnarray}
&~&(\psi_t ^\mu -i\eta \; 
S_{(t)\nu} ^\mu \; \tilde \psi_{-t} ^\nu)
|B_\psi , \eta \rangle=0. 
\label{eq:2.13}
\end{eqnarray}
Again by making use of the coherent state method,
the fermionic boundary state takes the feature
\begin{equation}
|B_\psi  ,\eta \rangle=\prod _{t}
[{\rm det} M_{(t)}] \exp \left[ i\eta \sum _{t}
(\psi_{-t} ^\mu S_{(t)\mu \nu} \; \tilde \psi_{-t} ^\nu) \right]
|0\rangle. 
\label{eq:2.14}
\end{equation}

The total boundary state is given by
\begin{equation}
|B,\eta \rangle_{\rm NS}=|B_x\rangle \otimes |B _\psi,
\eta \rangle_{\rm NS} \otimes |B_{\rm gh}\rangle \otimes
|B_{\rm sgh}, \eta \rangle_{\rm NS},
\label{eq:2.15}
\end{equation}
where the contribution of the superconformal ghosts
is given by 
\begin{eqnarray}
|B_{\rm sgh},\eta \rangle_{\rm NS}=\exp \left[ i\eta
\sum_{t=1/2} ^\infty \left(\gamma_{-t} \tilde \beta_{-t}
-\beta_{-t} \tilde \gamma_{-t}\right) \right]
|P=-1\rangle |\tilde P=-1 \rangle. 
\label{eq:2.16}
\end{eqnarray}
By employing the GSO projection the applicable boundary state is
written as a combination of the total boundary states 
with $\eta=\pm 1$,
\begin{eqnarray}
|B \rangle_{\rm NS}&=&\frac{1}{2}\left( |B ,+ \rangle_{\rm NS}
-|B ,-\rangle_{\rm NS}\right). 
\label{eq:2.17}
\end{eqnarray}

\subsection{The interaction in the NS-NS sector}

For computing the LREE we
need the partition function. Hence, we first introduce
the interaction amplitude between two identical and 
parallel D$p$-branes. The branes have been dressed by the
fields, and they have tangential dynamics. One can obtain
this amplitude from the overlap of the GSO-projected 
boundary states, associated with the two D$p$-branes,
via the propagator ``$D$'' of the exchanged closed string
\begin{eqnarray}
\mathcal{A}&=&\langle B_1|D|B_2\rangle,
\nonumber\\
D&=&2 \alpha' \int_0^\infty {\rm d}t\;e^{-tH},  
\label{eq:2.18}
\end{eqnarray}
in which ``$H$'' stands for the total Hamiltonian of the
propagating closed superstring. It
consists of the matter and ghost parts.
Therefore, one receives
\begin{eqnarray}
\mathcal A_{\rm NS-NS}&=& \frac{T_p ^2 V_{p+1} \alpha'}
{4(2\pi)^{9-p}}
\frac{1}{\sqrt{{\rm det} (U_1/4\pi \alpha'){\rm det}
(U_2/4\pi \alpha')}}
\prod_{m=1} ^\infty \frac{{\rm det} [M^\dagger _{(m-1/2)1}
M_{(m-1/2)2}]}{{\rm det}[M^\dagger _{(m)1}M_{(m)2}]}
\nonumber\\
&\times& \int^\infty _0 dt \;
\bigg\{ \left(\sqrt {\frac{1}{\alpha' t}}\right)^{9-p}
\exp \left(-\frac{1}{4\pi \alpha' t}
\sum_{i=p+1}^{9} (y_2^i-y_1^i ) ^2 \right)
\nonumber\\[10 pt]
&\times&\frac{1}{q}\bigg( \prod_{m=1}^\infty
\left[ \left(\frac{1+q^{2m-1}}{1-q^{2m}} \right)^{7-p}\;
\frac{{\rm det}(\textbf 1+Q^\dagger _{(m-1/2)1}Q_{(m-1/2)2} \;
q^{2m-1})}{{\rm det}(\textbf 1-Q^\dagger _{(m)1}Q_{(m)2} \;
q^{2m})}\right]
\nonumber\\[10 pt]
&-&\prod_{m=1}^{\infty}
\left[ \left( \frac{1-q^{2m-1}}{1-q^{2m}}\right)^{7-p}\;
\frac{{\rm det}(\textbf 1-Q^\dagger _{(m-1/2)1}Q_{(m-1/2)2} \;
q^{2m-1})}{{\rm det}(\textbf 1-Q^\dagger _{(m)1}Q_{(m)2} \;
q^{2m})}\right] \bigg)\bigg\}, 
\label{eq:2.19}
\end{eqnarray}
where $q=e^{-2\pi t}$, and $V_{p+1}$ indicates
the D$p$-brane worldvolume.
The two factors in the second line 
originate from the zero-modes and the factor $q^{-1}$ in the 
third line is related to the zero-point 
energy. For the first factors 
inside the infinite products we have
the power $7-p =[10-(p+1)]-2$, where
$10-(p+1)$ and $-2$ correspond to the contributions by 
the Dirichlet oscillators and ghosts-superghosts,
respectively. The numerators determinants
represent the contributions 
of the fermions Neumann oscillators, and 
that in the denominators is associated with 
the Neumann oscillators of the bosons.

The integer (half-integer)
modes exhibit the bosons (fermions) contribution.
The tension of a non-BPS brane includes an extra $\sqrt{2}$
factor, which in the above amplitude it has been considered.

\section{The LREE corresponding to the unstable
dressed-dynamical D$p$-brane}

Imagine a bipartite system which comprises only two 
subsystems A and B. Let the pure
state of the composite system be $|\psi \rangle$. Thus,
the density operator, associated with this state, is
defined by $\rho=|\psi \rangle \langle \psi |$.
The conservation of probability requires that ${\rm Tr}\rho=1$.
The reduced density operator due to the subsystem A
is defined as $\rho_{\rm A}={\rm Tr_B} \rho$, where the
${\rm Tr_B}$ represents the partial trace with respect to
the subsystem B.

The entanglement and R\'{e}nyi entropies are the 
most desirable tools among the other quantities for measuring
entanglement. The first quantity
can be obtained by the von Neumann formula
$S=-{\rm Tr}(\rho_{\rm A} \ln \rho_{\rm A})$ \cite{45}, while the
second one is derived from
$S_n=\frac{1}{1-n} \ln {\rm Tr}\rho_{\rm A} ^n$,
where $n\geq 0$ and $n\neq 1$. By taking the special limit, i.e.
$n \to 1$, the R\'{e}nyi entropy tends to 
the entanglement entropy \cite{46}.

\subsection{The density operator of the system}

The Hilbert space of closed superstring theory has the
factorized form $\mathcal{H}=\mathcal{H}_{\rm L}\otimes
\mathcal{H}_{\rm R}$. 
The left- and right-moving oscillating modes
of closed superstring form the bases of the 
subsystems ``L'' and ``R''.
For receiving the physical Hilbert space
we should exert the Virasoro constraints. 
Precisely, a general state of
closed superstring is given
by $|\psi \rangle=|\psi\rangle_{\rm L}
\otimes |\psi \rangle_{\rm R}$, where
\begin{eqnarray}
|\psi\rangle_{\rm L}&=&\prod_{k=1}^\infty \prod_{t}
\frac{1}{\sqrt{n_k !}}
\left(\frac{\alpha^{\mu_k}_{-k}}{\sqrt{k}}\right)^{n_k}
\left(\psi^{\mu_t}_{-t}\right)^{n_t}|0\rangle,
\nonumber\\
|\psi\rangle_{\rm R}&=&\prod_{k=1}^\infty \prod_{t}
\frac{1}{\sqrt{m_k !}}
\left(\frac{\tilde{\alpha}^{\nu_k}_{-k}}{\sqrt{k}}\right)^{m_k}
\left(\tilde \psi^{\nu_t}_{-t}\right)^{m_t}|0\rangle,
\nonumber\
\end{eqnarray}
where for the NS-NS sector the mode numbers ``$t$''
are positive half integers.
Since $\psi_{-t}^\mu$ and $\tilde \psi_{-t}^\nu$
are Grassmannian variables we have
$m_t,n_t \in\{0,1\}$.
The sets $\{n_t ,n_k|k\in \mathbb{N}\}$
and $\{m_t , m_k|k\in \mathbb{N}\}$ are independent
up to the condition 
\begin{eqnarray}
\sum_{k=1} ^\infty kn_k+\sum_t tn_t=
\sum_{k=1} ^\infty km_k+\sum_t tm_t. 
\nonumber\
\end{eqnarray}
The quantity in the left-hand side (right-hand side)
represents the total mode number, i.e.,
the summation of all mode numbers 
in the state $|\psi\rangle_{\rm L}$ ($|\psi\rangle_{\rm R}$). 
Thus, the Virasoro conditions at most 
impose only the equality of the total 
mode numbers of the states $|\psi\rangle_{\rm L}$ 
and $|\psi\rangle_{\rm R}$.
This condition weakly relates the left- and right-moving
string modes. Hence, the left- and right-sectors essentially
remain independent. Therefore, the
physical Hilbert space possesses the factorized form.

The boundary state, which is a coherent state of closed
superstring, is also decomposed to the left- and right-moving
modes by the Schmidt decomposition method
\cite{47}, \cite{48}.
In other words, the expansion of the exponential parts of
Eqs. (\ref{eq:2.6}) and (\ref{eq:2.14}) gives a series
which manifestly illustrates entanglement between the two 
parts of the Hilbert space.
Hence, similar to the non-geometric
prescription of Refs. \cite{8} and \cite{9},
we take the GSO-projected boundary state as the
composite system and
the left- and right-moving modes of closed
superstring as its subsystems.

The density operator, corresponding to 
a given boundary state, might
be considered as $\rho=|B\rangle \langle B|$. In fact,
the inner product $\langle B|B\rangle$ is divergent.
To see this, according to Eq. (\ref{eq:2.18}), in the
amplitude (\ref{eq:2.19}) remove the integral over 
``$t$'' and apply $t\to 0$. A consequence of this
divergence is violation of the condition ${\rm Tr}\rho=1$.
Thus, we consider the regularized state 
$|\mathcal B\rangle=(e^{-\epsilon H}/
\sqrt{\mathcal N_B})|B\rangle_{\rm NS-NS}$,
where $\epsilon$ is a finite correlation length.
Hence, the density operator is defined as
\begin{equation}
\rho = \frac{1}{\mathcal N_B}
\left(e^{-\epsilon H}
{|B\rangle _{\rm NS-NS}}\right)\;
\left({_{\rm NS-NS}\langle B|}e^{-\epsilon H}\right),
\label{eq:3.1}
\end{equation}
where the normalization factor $\mathcal N_B $
is fixed by the probability conservation condition 
${\rm Tr}\rho = 1$. 
After taking the trace of the density operator
over the closed superstring states
and applying ${\rm Tr}\rho = 1$, we obtain the 
normalization factor equal to the 
partition function 
$\mathcal N_B = Z_{\rm NS-NS}(2\epsilon)$.

In the paper \cite{5} there are two regularization 
approaches, which are corresponding to the boundary 
state and Ishibashi states. Each approach possesses 
its own normalization factor. As it has been shown 
in \cite{5} the regularization of the Ishibashi states
can correctly recover the spatial topological 
entanglement entropy for Chern-Simons theories 
while the first approach of regularization cannot 
recover it. However, unlike the topological 
theories, e.g. the Chern-Simons theories, our 
action does not represent a topological theory. 
That is, we don't have a topological sector, and hence, 
there is no any topological entanglement entropy. 
Thus, we don't normalize the Ishibashi states individually. 
Therefore, for the regularization we applied 
only the first approach. 

An interpretation of the numerator of $\rho$ 
is that a closed superstring propagates 
for the time $t=\epsilon$, then it is
absorbed by a D-brane. It is immediately emitted by an
identical D-brane and again propagates for the 
duration $t=\epsilon$. However, the interpretation 
of the partition function in the denominator
of (\ref{eq:3.1}), i.e. $Z_{\rm NS-NS}(2\epsilon)$, is that 
a closed superstring is emitted by a D-brane, then 
it propagates for the time $t=2\epsilon$ 
and then it is absorbed by an identical  
D-brane. 

The partition function can be conveniently 
extracted from the amplitude (\ref{eq:2.19}) as 
in the following
\begin{eqnarray}
Z_{\rm NS-NS}(2\epsilon)&=&
{_{\rm NS-NS}\langle B|}e^{-2\epsilon H}|B\rangle_{\rm NS-NS}
\nonumber\\[10pt]
 &=&\frac{T_p ^2\; V_{p+1}}
{2(2\pi)^{9-p}} \;
\frac{1}{{\rm det} (U/8\pi)}
\prod_{m=1} ^\infty \frac{|{\rm det} M _{(m-1/2)}|^2}
{|{\rm det}M _{(m)}|^2}
\left(\sqrt {\frac{1}{4 \epsilon}}\right)^{9-p}
\nonumber\\[10pt]
&\times&\frac{1}{q}\bigg( \prod_{m=1}^\infty
\left[ \left(\frac{1+q^{2m-1}}{1-q^{2m}} \right)^{7-p}
\frac{{\rm det}(\textbf 1+Q^\dagger _{(m-1/2)}Q_{(m-1/2)} \;
q^{2m-1})}{{\rm det}(\textbf 1-Q^\dagger _{(m)}Q_{(m)} \;
q^{2m})}\right]
\nonumber\\[10pt]
&-&\prod_{m=1}^{\infty}
\left[ \left( \frac{1-q^{2m-1}}{1-q^{2m}}\right)^{7-p}
\frac{{\rm det}(\textbf 1-Q^\dagger _{(m-1/2)}Q_{(m-1/2)} \;
q^{2m-1})}{{\rm det}(\textbf 1-Q^\dagger _{(m)}Q_{(m)} \;
q^{2m})}\right] \bigg). 
\label{eq:3.2}
\end{eqnarray}
Since we have identical branes in the same position,
the indices 1 and 2 and also the $y$-dependence 
have been omitted. Similar to the stringy literature,
in which for simplification various numeric 
values are chosen for the slope $\alpha'$
\cite{8, 9, 49},
we have selected the choice $\alpha'=2$.

\subsection{The associated LREE to the setup}

The first step for computing the LREE of our
setup is the calculation of the R\'{e}nyi entropy.
Accordingly, we need to find ${\rm Tr}\rho_{\rm L}^n$,
where the reduced density operator $\rho_{\rm L}$ is
derived via the trace over the right-moving oscillators.
We utilize the replica trick, which for the real ``$n$''
gives
\begin{equation}
{\rm Tr}\rho_{\rm L}^n\; \sim
\frac{Z_{\rm NS-NS}(2n\epsilon)}{Z_{\rm NS-NS}^n (2 \epsilon )}
\equiv \frac{Z_{n \; {\rm NS-NS}}({\rm L})}
{Z_{\rm NS-NS}^n}~. 
\label{eq:3.3}
\end{equation}
The quantity $Z_{n \; {\rm NS-NS}}$ is called the 
``replicated partition function''.

Since there are various approaches to sum over the spin 
structure ($\eta$) and momentum, there are different 
ways to acquire the replicated partition function and 
replicated normalization constant \cite{9}.
Explicitly, if we first sum
over $\eta$ and then we do the replication,
the spin structure of each copy will be disconnected from
the other copies. This case is called the \textit{uncorrelated}
spin structure. Another possibility is that:
at first replicate each spin
structure separately and then compute sum over 
them. This case is
called the \textit{correlated} spin structure.
In the same way, the
uncorrelated and correlated momentum are constructed by       
integrating over the momenta before and after the 
replication, respectively. Besides, 
if the normalization constant
$K_p^{1/2}$ (see Eq. (\ref{eq:3.5}))
is raised to the power $n$, through 
the replication process,
we call it \textit{replicated} 
normalization constant. Otherwise,
it will be called the \textit{unreplicated} 
normalization constant.

In fact, all of the above possibilities can be studied. 
However, here we
choose only one of them which is invariant under the
open-closed string duality. This reliable case possesses the
unreplicated normalization constant, the correlated
momentum and the correlated spin structure.
For an NS-NS brane we have 
\begin{eqnarray}
\int_{0}^\infty {\rm d}l \;_{\rm NS-NS}\langle B,
\eta|e^{-lH_c}|B,\eta \rangle_{\rm NS-NS}
&=&{\mathcal N}^2 \int_{0}^\infty {\rm d}l
\left ( \frac{1}{l} \right) ^{\frac{9-p}{2}}
\frac{f_3^8 (q)}{f_1^8 (q)}
\nonumber\\
&=&{\mathcal N}^2 \; \frac{32 (2\pi)^{p+1}}{V_{p+1}}
\int_{0}^\infty \frac{{\rm d}t}{2t} {\rm Tr}_{\rm NS}
\left[ e^{-tH_o} \right],
\nonumber\\
\int_{0}^\infty {\rm d}l \;_{\rm NS-NS}\langle B,
\eta|e^{-lH_c}|B,-\eta \rangle_{\rm NS-NS}
&=&{\mathcal N}^2 \int_{0}^\infty {\rm d}l
\left ( \frac{1}{l} \right) ^{\frac{9-p}{2}}
\frac{f_4^8 (q)}{f_1^8 (q)}
\nonumber\\
&=&{\mathcal N}^2 \; \frac{32 (2\pi)^{p+1}}{V_{p+1}}
\int_{0}^\infty \frac{{\rm d}t}{2t} {\rm Tr}_{\rm R}
\left[ e^{-tH_o} \right],
\nonumber\
\end{eqnarray}
where the integral variables $l$ 
and $t$ exhibit the length of 
the cylinder in closed string 
channel and the circumference of 
the cylinder in the open string 
channel, respectively. 
For a non-BPS brane the normalization constant is
\begin{eqnarray}
{\mathcal N}^2_{\rm non-BPS}=
\frac{V_{p+1}}{64(2\pi)^{p+1}}.
\nonumber\
\end{eqnarray}
The replicated partition function 
with the correlated momentum 
gives rise to the factor $(1/nl)^{{(9-p)}/2}$, 
while $Z_n$ with 
uncorrelated momentum leads to the factor 
$(1/l)^{{n(9-p)}/2}$, which is not 
invariant under the modular transformation. 
Besides, the correlated spin 
structure leads to a factor 2, while the 
uncorrelated spin structure introduces the factor $2^{2n-1}$. 
In addition, the unreplicated normalization ${\mathcal N}^2$ 
is chosen instead of the replicated 
normalization ${\mathcal N}^{2n}$. These imply that
to satisfy the open-closed duality we have to apply 
the replicated partition function 
with the correlated momentum, 
unreplicated normalization constant, and the
correlated spin structure.

As $\epsilon$ tends to zero the quantity 
$q=e^{-4\pi \epsilon}$ does not vanish. Therefore,
we apply the transformation $4\epsilon \to 1/4\epsilon$
to go to the open string channel. Here, we 
work with the quantity $\tilde q=\exp{(-\frac{\pi}{4\epsilon})}$
which in the limit $\epsilon \to 0$ tends to zero.
Thus, we can expand Eq. (\ref{eq:3.3}) for small $\tilde q$
as in the following
\begin{eqnarray}
\frac{Z_{n \; {\rm NS-NS}}}{Z_{\rm NS-NS}^n}
&\approx&2^{1-n}\;K^{1-n}_p
\left( \left(2\sqrt{\epsilon}\right)^{1-n}\sqrt{n}\right)^{9-p}
\exp \left[\frac{\pi}{4\epsilon}\left(\frac{1}{n}-n\right)\right]
\nonumber\\[10pt]
&\times&\prod_{m=1}^\infty 2^{1-n}\;
C_{(m-1/2)}^{1-n}\bigg\{\tilde q^{\frac{2m-1}{n}-n(2m-1)}
+C_{(m)}\;\tilde q^{\frac{4m-1}{n}-n(2m-1)}
\nonumber\\[10pt]
&-&n\;C_{(m)}\;\tilde q^{\frac{2m-1}{n}-n(2m-1)
+2m}-n\;C^2_{(m)}\;
\tilde q^{\frac{4m-1}{n}-n(2m-1)+2m}
\nonumber\\[10pt]
&+&\frac{n(n+1)}{2}\;C^2_{(m)}\;
\tilde q^{\frac{2m-1}{n}-n(2m-1)+4m}
+\mathcal O(\tilde q^{6m})\bigg\},
\label{eq:3.4}
\end{eqnarray}
where $K_p$, $C_{(m)}$ and $C_{(m-1/2)}$ are defined by 
\begin{eqnarray}
K_p&=& \frac{T_p ^2\; V_{p+1}}{2(2\pi)^{9-p}} \;
\frac{1}{{\rm det} (U/8\pi)}
\prod_{m=1} ^\infty \frac{|{\rm det} M _{(m-1/2)}|^2}
{|{\rm det}M _{(m)}|^2},
\nonumber\\[10pt]
C_{(t)}&=&{\rm Tr}\left(Q^\dagger_{(t)}Q_{(t)}\right)
+7-p\; . 
\label{eq:3.5}
\end{eqnarray}
The index ``$t$'' is a positive integer ``$m$''
or a positive half-integer ``$m-1/2$''.

Now for obtaining the LREE we should take the 
limit $n\to 1$ of the R\'{e}nyi entropy, which yields
\begin{eqnarray}
S^{(p)}_{\rm LREE}&\approx&\frac{1}{2} 
\ln{2}+\ln K_p+\frac{9-p}{2}
\left(2\ln 2+\ln \epsilon -1\right)+\frac{\pi}{3\epsilon}
\nonumber\\[10pt]
&+&\sum_{m=1}^\infty
\bigg\{ \ln C_{(m-1/2)}
+C_{(m)}\left(1-\frac{m\pi}{2\epsilon}\right)
e^{-{m\pi}/{2\epsilon}}
\nonumber\\[10pt]
&-&\frac{1}{2}C_{(m)}^2 \left(1-\frac{m\pi}{\epsilon}\right)
e^{-{m\pi}/{\epsilon}}
+\mathcal O(\exp(-3m\pi/2\epsilon))\bigg\}. 
\label{eq:3.6}
\end{eqnarray}
The first term comes from the sum over the spin structure
and a contribution from the oscillators. The second term
shows the boundary entropy of the brane, the third term 
originates from the zero-modes, 
and the rest terms are regarding 
to the contributions of the oscillators and conformal 
ghosts. The parameters of the setup have been appeared
in $K_p$, $C_{(m)}$ and $C_{(m-1/2)}$. Besides,
the mode dependence of the LREE is a consequence of
the presence of the tachyonic field.

\subsection{The LREE and the thermodynamic entropy}

To investigate the thermal properties of our system
we can associate a temperature to it.
This temperature is proportional to the
inverse of the correlation length, i.e. $\beta=2\epsilon$.
Applying the definition of the thermodynamic entropy
and using the partition function (\ref{eq:3.2}), in 
the high temperature limit of the 
system $\epsilon \to 0$, we find
\begin{eqnarray}
S_{\rm thermal}&=&\beta^2 \; \frac{\partial}{\partial \beta}
\left(-\frac{1}{\beta}\ln Z_{\rm NS-NS} \right)
\nonumber\\[10pt]
&\approx&\frac{1}{2}\ln 2+\ln K_p
+\frac{9-p}{2}
\left(\ln 2\beta-1\right)+\frac{2\pi}{3\beta}
\nonumber\\
&+&\sum_{m=1}^\infty \bigg\{\ln C_{(m-1/2)}
+C_{(m)}\left(1-\frac{m\pi}{\beta}\right)
e^{-{m\pi}/{\beta}}
\nonumber\\[10pt]
&-&\frac{1}{2}\; C_{(m)}^2
\left(1-\frac{2m\pi}{\beta}\right)
e^{-{2m\pi}/{\beta}}+\mathcal O(\exp(-3m\pi/\beta))\bigg\}.
\label{eq:3.7}
\end{eqnarray}
We observe that the thermodynamic entropy of the 
system exactly is equivalent to its LREE.
This similarity between the thermal and entanglement
entropies also has been obtained in the literature,
e.g., see Refs. \cite{38}-\cite{42}.

Since the constants $\{C_{(m)}|m\in \mathbb{N}\}$ 
depend on the mode numbers calculation of the 
summation of the series in Eq. (\ref{eq:3.7}) is
very complicated. 
Therefore, we don't have an explicit form of the 
entropy function $S_{\rm thermal}(T)$, 
in which $\beta =1/T$. 
Hence, the phase transition of the corresponding system 
is not clear.
\section{Condensing the tachyon}

\subsection{Evolution of the LREE under the 
tachyon condensation}

Presence of an open string tachyon
on a D-brane drastically makes it unstable. 
Through the tachyon condensation process 
the D-brane collapses, i.e., it looses 
some of its directions.
Ultimately, one receives the closed string vacuum 
or at most an intermediate stable D-brane \cite{29, 30}.
Under the tachyon condensation at least one of the
elements of the tachyon matrix $U_{\alpha \beta}$
tends to infinity. For instance, if we apply 
$U_{pp}\to \infty$ the condensation occurs 
in the $x^p$-direction.

Before imposing the condensation on the tachyon 
we compute the LREE in the large value of the tachyon matrix,
that is $U\gg 2(\Delta-\mathcal F)$. This tachyon matrix 
accompanied by the conditions (\ref{eq:2.8}) yield 
\begin{eqnarray}
{\tilde S}^{(p)}_{\rm LREE}&\approx& \ln 2+\ln K_p+\frac{9-p}{2}
\left(2 \ln 2+\ln \epsilon -1\right)
+\frac{\pi}{3\epsilon}
\nonumber\\[10pt]
&+&\sum_{m=1}^\infty \bigg\{ \ln H_{(m-1/2)}
+H_{(m)}
\left(1-\frac{m\pi}{2\epsilon}\right)
e^{-{m\pi}/{2\epsilon}}
\nonumber\\[10pt]
&-&\frac{1}{2}H_{(m)}^2
\left(1-\frac{m\pi}{\epsilon}\right)
e^{-{m\pi}/{\epsilon}}+\mathcal O(\exp(-3m\pi/2\epsilon))
\bigg\}, 
\label{eq:4.1}
\end{eqnarray}
up to the order $\mathcal{O}(U^{-3})$,
where we defined
\begin{equation}
H_{(t)}=8-512\;t^2
{\rm Tr} \left(\omega^2 U^{-2}\right), 
\label{eq:4.2}
\end{equation}
The index ``$t$'' is a positive integer ``$m$''
or a positive half-integer ``$m-1/2$''.

Now, suppose that the tachyon is condensed only in the
$x^p$-direction of the brane. In this case, one finds
\begin{eqnarray}
\lim_{U_{pp} \to \infty} \ln K_p
&=&\ln K_{p-1}+\ln \left(\frac{\pi L_p}{{\bar U}_{pp}}\right), 
\label{eq:4.3}
\end{eqnarray}
in which the infinite value of $U_{pp}$ 
was called ${\bar U}_{pp}$, 
and $L_p$ is the infinite length of the brane in the 
$x^p$-direction. For acquiring this result,
the trusty relation $T_p=T_{p-1}/(2\pi \sqrt{\alpha'})$  
and the regularization schemes 
$\prod_{n=1}^\infty n \to \sqrt{2\pi}$
and $\prod_{n=1}^\infty (2n-1) \to \sqrt{2}$ have
been exerted. The third phrase of Eq. (\ref{eq:4.1})
can be rephrased as
\begin{eqnarray}
\frac{9-p}{2} \;(2\ln 2+\ln \epsilon-1)
&=& \frac{9-(p-1)}{2}\;(2\ln 2 +\ln \epsilon-1)
\nonumber\\
&-&\frac{1}{2}\left(2\ln 2+\ln \epsilon -1\right) .
\label{eq:4.4}
\end{eqnarray}
By taking the limit $U_{pp} \to \infty$, the factor
${\rm Tr}\left( \omega U^{-2}\right)$ reduces to
${\rm Tr}\left( \omega U^{-2}\right)'$, where the
prime indicates a $p\times p$ matrix. Accordingly,
under the tachyon condensation experience the LREE 
finds the form  
\begin{eqnarray}
\lim_{U_{pp} \to \infty}{\tilde S}^{(p)}_{\rm LREE}
={\tilde S}^{(p-1)}_{\rm LREE}+\lambda , 
\label{eq:4.5}
\end{eqnarray}
\begin{eqnarray}
\lambda &\equiv & \ln \left(\frac{\pi L_p}
{2{\bar U}_{pp}}\right)
-\frac{1}{2}(\ln \epsilon-1). 
\label{eq:4.6}
\end{eqnarray}
In fact, 
when the tachyon condensation acts on one 
direction of an unstable D$p$-brane,
it collapses to a D$(p-1)$-brane \cite{30}. Here, 
the associated LREE with the D$(p-1)$-brane 
is exactly given by ${\tilde S}^{(p-1)}_{\rm LREE}$.
The infinite parameters $L_p$ and ${\bar U}_{pp}$ can be
accurately adjusted such that their ratio to be a finite
value. 

The extra contribution to the entropy, 
i.e. $\lambda$, can be
interpreted as the entropy of the released closed 
superstrings via the collapse of the D$p$-brane.
In comparison with the bosonic case 
\cite{28}, the extra entropy
$\lambda$ has reduced by $-\ln (2{\bar U}_{pp})$, 
which can 
be interpreted as reduction of superstring 
radiation during the collapse of the brane.
For example, consider the case that the total 
entropies of the bosonic and superstring systems,
after tachyon condensation, are 
equal. Then, the inequality 
$\lambda_{\rm bosonic} > \lambda_{\rm superstring}$ 
induces the following inequality  
\begin{eqnarray}
\left({\tilde S}^{(p-1)}_{\rm LREE}\right)
_{\rm superstring}
> \left({\tilde S}^{(p-1)}_{\rm LREE}\right)
_{\rm bosonic\;string}. 
\nonumber
\end{eqnarray}
Thus, one may deduce that under the tachyon condensation 
the resultant D$(p-1)$-brane in the superstring theory 
is more stable than that in the bosonic string theory.

\subsection{The second law of thermodynamics for the LREE}

The thermal and entanglement entropies
have some close connections \cite{38}-\cite{42}. 
For instance, in Refs. \cite{38}-\cite{40} it has been
demonstrated that the entanglement entropy obeys relations
which are 
similar to the laws of thermodynamics. In Sec. (3.3)
we proved that the LREE and thermal entropy of
our setup possess an identical feature. This 
similarity stimulated us to check the second law of
thermodynamics for the LREE under the tachyon condensation
process.

Now we compare the LREE of our
initial state, which is the D$p$-brane,
with that of the final state, i.e. the resultant 
D$(p-1)$-brane and the released closed superstrings. 
Thus, we have
\begin{eqnarray}
S_{\rm initial}&=&{\tilde S}^{(p)}_{\rm LREE}\;,
\nonumber\\
S_{\rm final}&=&\lim_{U_{pp} \to \infty}
{\tilde S}^{(p)}_{\rm LREE}={\tilde S}^{(p-1)}_{\rm LREE}
+\lambda .
\label{eq:4.7}
\end{eqnarray}
The second law of thermodynamics implies that,
the entropy should be increased during the process.
Therefore, we should check the inequality
$S_{\rm final}-S_{\rm initial} > 0$,
\begin{eqnarray}
{\tilde S}^{(p-1)}_{\rm LREE}+\lambda
-{\tilde S}^{(p)}_{\rm LREE}
&=&\ln \left(\frac{\pi}{2{\bar U}_{pp}}\right)
-\ln\left(\frac{\det U'}{\det U}\right)
-\sum^\infty_{m=1}\bigg \{ 2\ln \left(\frac{\det
{M'}_{(m)}}{\det M_{(m)}} \frac{\det
{M}_{(m-1/2)}}{\det M'_{(m-1/2)}}\right)
\nonumber\\[10pt]
&+&\ln \left(\frac{H_{(m-1/2)}}{H'_{(m-1/2)}}\right)
+\left(H_{(m)}-H'_{(m)}\right)
\left(1-\frac{m\pi}{2\epsilon}\right)
e^{-{m\pi}/{2\epsilon}}
\nonumber\\[10pt]
&-& \frac{1}{2}\left( H^2_{(m)}-{H'}^2_{(m)}\right)
\left(1-\frac{m\pi}{\epsilon}\right)
e^{-{m\pi}/{\epsilon}}\bigg \}. 
\label{eq:4.8}
\end{eqnarray}
The primes represent the $p\times p$ matrices and
$H_{(m)}$ was defined by Eq. (\ref{eq:4.2}).
There are many parameters, i.e. the
various matrix elements, which control the
value of this difference.
The minimal condition for positivity of 
(\ref{eq:4.8}) is given by 
\begin{eqnarray}
\frac{\det U}{\det U'}
\prod^\infty_{m=1}\left[\left(\frac{\det M_{(m)}
\det {M'}_{(m-1/2)}}
{\det M'_{(m)}\det M_{(m-1/2)}}
\right)^2 \;\frac{H_{(m-1/2)}}{H'_{(m-1/2)}} \right]>
\frac{2{\bar U}_{pp}}{\pi}, 
\label{eq:4.9}
\end{eqnarray}
up to the leading order. According to the following formula 
\begin{eqnarray}
\cos \theta =  \prod^\infty_{m=1}\left[ 1-
\frac{\theta^2}{(m-1/2)^2\pi^2}\right],
\nonumber
\end{eqnarray}
we can write 
\begin{eqnarray}
\prod^\infty_{m=1}\frac{H_{(m-1/2)}}{H'_{(m-1/2)}} 
&=&\frac{\cos \phi}{\cos \phi'}
\prod^\infty_{m=1}\frac{{\rm Tr}(\omega^2 U^{-2})}
{{\rm Tr}(\omega^2 U^{-2})'}
\nonumber\\
&=&\frac{\cos \phi}{\cos \phi'}
\left( \frac{{\rm Tr}(\omega^2 U^{-2})}
{{\rm Tr}(\omega^2 U^{-2})'}\right)^N,
\label{eq:4.10}
\end{eqnarray}
where $N=\sum^\infty_{m=1}1$, and 
the angle $\phi$ has the definition 
\begin{eqnarray}
\phi = \frac{\pi}{8\sqrt{{\rm Tr}(\omega^2 U^{-2})}}\;.
\nonumber
\end{eqnarray}
Now we impose an additional condition 
\begin{eqnarray}
R \equiv \frac{{\rm Tr}(\omega^2 U^{-2})}
{{\rm Tr}(\omega^2 U^{-2})' } > 1.
\label{eq:4.11}
\end{eqnarray}
This inequality inspires that the second 
factor in the RHS of Eq. (\ref{eq:4.10}) 
is infinite. In fact, the infinities in the 
LHS and RHS of (\ref{eq:4.9}) completely are
independent. However, the value of the quantity 
$R$ depends on all matrix elements of the 
matrices $U$ and $\omega$. By adjusting the 
parameters 
$\{U_{\alpha \beta},\omega_{\alpha \beta}|
\alpha ,\beta =0,1, \cdots ,p\}$
we can receive a large
value for $R$ such that the 
infinity in the LHS of (\ref{eq:4.9})
to be dominant to ${\bar U}_{pp}$, and the ratio 
$R^N/{\bar U}_{pp}$
to be fixed. Finally, these conditions
reliably confirm the preservation of the second law of 
thermodynamics for the LREE of the setup.

\section{Conclusions}

In the context of the type IIA/IIB superstring theories
we investigated the left-right entanglement entropy 
of a non-BPS unstable D$p$-brane. 
The brane has tangential dynamics. Besides, they
have been dressed by the $U(1)$ gauge potential, 
the anti-symmetric tensor
field and the open string tachyon field.
For achieving this, the boundary state
formalism in the NS-NS sector was employed 
and the interaction amplitude
between two identical dynamical D$p$-branes 
with the foregoing fields was introduced.

The parameters of the
dynamics and background fields were entered into the LREE,
and hence, they generalized the form of the LREE.
Therefore, the value of the LREE can be 
accurately controlled by adjusting 
these parameters. Because of the 
presence of the tachyon field,
the closed string mode numbers  
drastically appeared in the LREE
through the infinite product and the series. However,
as we chose only the NS-NS sector,
both the integer and half-integer modes were entered.

Effect of the tachyon condensation on the LREE was also
studied. The LREE of the initial  
D$p$-brane was decomposed to the LREE of a new unstable 
dressed-dynamical D$(p-1)$-brane and 
an extra contribution which belongs to the 
emitted closed superstrings through the brane collapse.
In comparison with the bosonic case \cite{28}, 
the extra entropy has been reduced, which indicates 
a smaller amount of string radiation. This reveals that
after the tachyon condensation  
the resultant D-brane in the superstring theory is more stable
than its counterpart in the bosonic string theory.

Furthermore, we defined a temperature for our
system to derive the thermodynamic entropy via the 
partition function. We
found that the thermal entropy of the configuration 
exactly is equivalent to its LREE.
Similar equivalence relations have been demonstrated
in Refs. \cite{9, 27, 28}. The common properties
of the thermodynamic entropy and LREE 
motivated us to check the second law of thermodynamics 
for the LREE under the tachyon condensation process.
In fact, preservation
of the second law of thermodynamics for the 
LREE imposes two prominent conditions
among the parameters of the setup.


\end{document}